\documentclass[12pt]{article}
\usepackage[dvips]{epsfig}

\begin{document}
\begin{flushright}
OHSTPY-HEP-T-98-010 \\
hep-th/9805188
\end{flushright}
\vspace{20mm}
\begin{center}
{\LARGE On the Transition from Confinement to Screening in
   $\mbox{QCD}_{1+1}$ Coupled to Adjoint Fermions at Finite $N$}
\\
\vspace{20mm}
{\bf F.Antonuccio and S.Pinsky} \\
\vspace{10mm}
{\em Department of Physics,\\ The Ohio State University,\\ Columbus,
OH 43210}
\end{center}
\vspace{20mm}
\begin{abstract}

We consider SU($N$) $\mbox{QCD}_{1+1}$ 
coupled to massless adjoint Majorana fermions, where
$N$ is finite but arbitrary.
We examine the spectrum for various values of $N$,   
paying particular attention to the formation
of multi-particle states, which were
recently identified by Gross, Hashimoto and Klebanov
in the $N=\infty$ limit of the theory.
It is believed that in the limit of vanishing fermion
mass, there is a transition from confinement to
screening in which string-like states made out of adjoint 
fermion bits dissociate into stable constituent ``single particles''.
In this work, we provide numerical evidence that 
such a transition into stable constituent
particles occurs not only at large $N$, but for any finite 
value of $N$. In addition, we discuss
certain issues concerning the ``topological'' properties
exhibited by the DLCQ spectrum.

\end{abstract}
\newpage

\baselineskip .25in

\section{Introduction}

Solving for the non-perturbative properties of 
physically realistic quantum
gauge theories is typically an intractable problem. In order
to gain some   insights, however, a number of lower dimensional models have
been investigated in the large $N$ (or planar) approximation, with 
a plethora of examples emerging 
in the last few years (see reference \cite{bpp98} for an extensive review).

In this work, we will consider
the $1+1$ dimensional SU($N$) gauge theory of QCD coupled to
massless adjoint Majorana fermions, which 
is believed to exhibit the property of 
screening \cite{gkm96,ars97}.
A similar theory with complex adjoint fermions has also been considered
\cite{anp97,pin97}.
 It will be 
advantageous to quantize 
the theory on the light-cone, and to adopt the light-cone gauge.
It is then a straightforward task to
extract numerical bound state solutions via an application of 
Discrete Light-Cone Quantization (DLCQ) \cite{pb85}.

Various non-perturbative studies of this model already exist
in the literature \cite{bdk93,ku93,ks95},
but in more recent work
\cite{ghk97} by Gross, Hashimoto and Klebanov (hereafter `GHK'),
it was suggested that in the massless fermion limit,
the spectrum becomes continuous above a certain threshold. 
This was supported by the presence of states in the spectrum
that have a mass consistent with the dynamics of two freely 
interacting stable particles. In this context, identifying either
``single particle''  or ``multi-particle'' states
requires a careful analysis of the spectrum, since 
an analysis of the explicit Fock state content does not
distinguish these states unambiguously.

It is worth clarifying this last remark to avoid possible 
confusion. Firstly, since the limit $N = \infty$ is assumed in GHK, 
Fock states are {\em single} traces of fermion creation operators,
and correspond intuitively to single closed strings of adjoint fermion
bits. Multi-trace states (corresponding to multi-string states)
do not appear in the analysis, since
interactions with them are suppressed (the factor $1/N$ plays
the role of a string coupling constant). 
Naively, one would view single-trace states as ``single particles'' and
multi-trace states as ``multi-particles''. Although
the latter is expected to follow from the usual $1/N$ counting,
GHK have shown that there may be exceptions to the former; namely,
states that are superpositions of 
single-trace Fock states may exhibit masses that are consistent with
the dynamics of
two freely interacting stable particles. 

This property of the large $N$ spectrum was perhaps anticipated
by the work of Kutasov and Schwimmer \cite{ks95,kub97};
namely, the presence of multi-particle states in the $N=\infty$ theory
becomes manifest after an appropriate rearrangement of the Hilbert space
into Kac Moody current blocks.
At finite $N$, however,  $1/N$ interactions
prevent us from identifying obvious multi-particle candidates
in the Hilbert space\footnote{For large $N$,
multi-trace states are obvious candidates for multi-particles.}, 
and we can no longer appeal to 
the correspondence proposed in \cite{ks95} to argue for the
existence of multi-particle states. 
Nevertheless, the formation of multi-particles  
in the finite $N$ spectrum may be inferred from the screening
properties of the theory\cite{ghk97}, since the theory
screens for any finite $N$\cite{gkm96}. 
It is therefore of interest to determine
whether multi-particles actually
appear in the finite $N$ spectrum or not,
and we will devote ourselves towards answering this question
via a numerical study of the DLCQ bound state equations.
 We also attempt to shed light on
certain ``topological'' features exhibited by the DLCQ spectrum.

Of course, working at finite $N$ has a price; the Fock space now 
admits multi-trace states, and the complexity of the 
bound state problem is dramatically increased. Nevertheless, we find 
that the numerical bound state problem is still tractable 
provided the discretization of the light-cone momentum $P^+$
is not too fine, but fine enough to resolve certain features of the 
theory we are interested in.

The organization of the paper may be summarized as follows; in 
Section \ref{dlcqformulation} we
briefly review the $1+1$ SU($N$) gauge theory coupled to
adjoint Majorana fermions, which we formulate in light-cone coordinates.
We also discuss features of the DLCQ formulation at finite $N$ that
differ from the large $N$ formulation originally discussed in \cite{bdk93}.
In Section \ref{numericalresults} we
tabulate the results of our finite $N$ numerical analysis, and
compare spectra of candidate multi-particle states
with mass predictions given
by free-body kinematics. We conclude our investigation
with a summary and discussion in Section \ref{conclusions}.

\section{Light-Cone Quantization and DLCQ at Finite $N$}
\label{dlcqformulation}

The action for $\mbox{QCD}_{1+1}$
coupled to a single Majorana fermion transforming 
in the adjoint representation of U($N$) or SU($N$) is given 
by\footnote{We consider U($N$) as well as SU($N$), since 
it was shown in earlier work \cite{alp98ab} that the SU($N$)
spectrum may be obtained by solving for the U($N$) spectrum,
and then `factoring out' U(1) states. Numerically,
this method is considerably
more efficient than solving for the SU($N$) spectrum directly.} 
\begin{equation}
S=\int d^2 x \mbox{Tr}\left[ {\rm i} {\overline \Psi } \gamma^{\mu} D_{\mu}
    \Psi - m{\overline \Psi } \Psi - \frac{1}{4 g^2} F_{\mu \nu}F^{\mu \nu}
   \right],  
\end{equation}
where the covariant derivative is defined by 
$D_{\mu}\Psi = \partial_{\mu} \Psi + {\rm i}[A_{\mu},\Psi]$,
and the fermion field $\Psi = 2^{-1/4}\left( \begin{array}{c}
                                                   \psi \\
                                                   \chi \end{array}  
                                      \right)$ 
is a two component spinor, each component representing an $N \times N$
Hermitian matrix of Grassmann variables. These matrices are traceless
if the gauge group is SU($N$). The light-cone quantization of
this theory in the light-cone gauge $A_- = 0$ was carried out
in \cite{bdk93}, and we refer the reader to that source for
details and notation. In the present context, we simply note that the 
light-cone momentum $P^+$ 
and Hamiltonian $P^-$ may be expressed
in terms of Fourier oscillator modes as follows:
\begin{eqnarray}
\lefteqn{P^+  =  \int_0^{\infty} dk \hspace{1mm} k \hspace{1mm} 
 b^{\dagger}_{ij}(k)b_{ij}(k),} \label{pplus} \\
\lefteqn{ P^-  =  \frac{1}{2} m^2  \int_0^{\infty} \frac{dk}{k} 
\hspace{1mm} 
 b^{\dagger}_{ij}(k)b_{ij}(k) + \frac{g^2 N}{\pi}
  \int_0^{\infty} \frac{dk}{k} 
\hspace{1mm} D(k) \left(  b^{\dagger}_{ij}(k)b_{ij}(k) -
    \frac{1}{N}  b^{\dagger}_{ii}(k)b_{jj}(k) \right) } \nonumber \\
& & + \frac{g^2}{2 \pi} \int_0^{\infty} dk_1 dk_2 dk_3 dk_4 \left\{ \frac{}{}
   \delta(k_1+k_2-k_3-k_4) \left[ A(k_i)\cdot
    b^{\dagger}_{kj}(k_3)b^{\dagger}_{ji}(k_4)
    b_{kl}(k_1)b_{li}(k_2) \right. \right. \nonumber \\
& & \hspace{5mm} + \left. B(k_i) 
  \cdot b^{\dagger}_{ij}(k_3)b^{\dagger}_{kl}(k_4)
          b_{il}(k_1)b_{kj}(k_2) \right] + 
   \delta(k_1+k_2+k_3-k_4)\times \nonumber \\
& & \hspace{5mm} C(k_i) \cdot \left[ 
     b^{\dagger}_{kj}(k_4)b_{kl}(k_1)
     b_{li}(k_2)b^{\dagger}_{ij}(k_3) -
     b^{\dagger}_{kj}(k_1)b^{\dagger}_{jl}(k_2)
     b^{\dagger}_{li}(k_3)b_{ki}(k_4) \right]\left. \frac{}{} \right\},
\label{pminus} 
\end{eqnarray}  
where
\begin{eqnarray}
   A(k_i) & = & \frac{1}{(k_4-k_2)^2} - \frac{1}{(k_1+k_2)^2}, \\
   B(k_i) & = & \frac{1}{2}\left( \frac{1}{(k_1-k_4)^2} - 
                \frac{1}{(k_2-k_4)^2} \right), \\
   C(k_i) & = & \frac{1}{(k_2+k_3)^2} - \frac{1}{(k_1+k_2)^2}, \\
   D(k)   & = & \int_0^{k} dp \hspace{1mm} \frac{k}{(p-k)^2}.
\end{eqnarray}
For the gauge group U($N$), the fermionic oscillator modes above
satisfy the following anti-commutation relations 
\begin{equation}             
  \{ b_{ij}(k^+),b^{\dagger}_{lk}({\tilde k}^+) \}
     = \delta(k^+ - {\tilde k}^+)\delta_{il}\delta_{jk}.
\label{unrelations}
\end{equation}
If the gauge group is SU($N$) we need to adopt the following set of 
relations --
\begin{equation}             
  \{ b_{ij}(k^+),b^{\dagger}_{lk}({\tilde k}^+) \}
     = \delta(k^+ - {\tilde k}^+)(\delta_{il}\delta_{jk}
              - \frac{1}{N} \delta_{ij}\delta_{kl}),
\label{sunrelations}
\end{equation}
and discard the $\frac{1}{N}b^{\dagger}_{ii}(k)b_{jj}(k)$ term 
in $P^-$, since the SU($N$) fermion fields are traceless.

A number of comments are in order. If we compare the 
light-cone Hamiltonian $P^-$ of equation (\ref{pminus}) -- which
is valid for any $N$ -- to the large $N$ expression given
in \cite{bdk93}, 
we notice that there is 
an additional $bb \rightarrow bb$ 
term in (\ref{pminus})  with
amplitude $B(k_i)$. The color index structure of this 
term\footnote{Repeated indices are always summed from $1$ to $N$.}
implies that it is suppressed by $1/N$, 
since it splits a single-trace Fock state
into multi-trace states. Nevertheless, the
$Z_2$ symmetry $T:b_{ij} \rightarrow b_{ji}$ 
of the large $N$ theory \cite{ku93} is also 
manifest in the finite $N$ formulation,
since this additional term is easily shown to be invariant under $T$.

The $D$ term in equation (\ref{pminus})
is obtained by normal ordering quartic terms in $P^-$, and diverges
linearly. However, this divergence is 
entirely absorbed by the Coulomb divergence generated by
the $A$ term. So the theory is manifestly finite \cite{bdk93}.
We also note that the $D$ term used in the large $N$ expression for $P^-$
is unchanged if $N$ is finite -- the
$1/N$ contributions 
obtained from normal ordering
quartic terms in accordance with the SU($N$) relations
 (\ref{sunrelations}) simply sum to zero.

\noindent

\medskip

In order to implement the DLCQ formulation of
the bound state problem \cite{pb85}, we simply
restrict the light-cone momentum variables $k_i$
appearing in equations (\ref{pplus}),(\ref{pminus})
for $P^{\pm}$ to the following set of discretized 
momenta:  $\{\frac{P^+}{K},\frac{3P^+}{K},
\frac{5P^+}{K},\dots \}$; i.e. only {\em odd} positive
integer multiples of $P^+/K$ are allowed, which is equivalent
to imposing anti-periodic boundary conditions\footnote{
Periodic boundary conditions are also possible, but convergence
in numerical calculations is slower \cite{bdk93}.} for the fermion fields:
$\psi_{ij}(x^-) = -\psi_{ij}(x^- + 2\pi R)$.
The integer $K$
is called the {\em harmonic resolution},
and $1/K$ measures the coarseness of our discretization.
Physically, $1/K$ represents the smallest unit of longitudinal
momentum fraction allowed for each parton. As soon as we implement the
DLCQ procedure, which is specified unambiguously by the harmonic 
resolution $K$, the integrals appearing in the definitions
(\ref{pplus}),(\ref{pminus}) for $P^{\pm}$ are replaced by
finite sums, and the eigen-equation
$2P^+ P^- |\Psi\rangle = M^2  |\Psi\rangle$ is reduced to a finite 
matrix problem. Continuum values are obtained
by extrapolating results to the $K=\infty$ limit.
Typically, a computer program is used to generate 
and diagonalize the DLCQ matrix to solve for the mass eigenvalues $M^2$.
In the present work, we are able to perform numerical diagonalizations
for values of $K$ in the range $3 \leq K \leq 16$ with the help
of Mathematica and a desktop PC. 
At $K=16$, the DLCQ matrix has dimensions $375 \times 375$.
A similar finite $N$ analysis 
of a two dimensional supersymmetric matrix model was performed recently 
by the authors \cite{alp98ab}.

\section{DLCQ Bound State Solutions}
\label{numericalresults}

In this section we present the results of our 
numerical diagonalizations of the DLCQ matrix for $M^2 = 2P^+P^-$
for values of $K$ in the range $3 \leq K \leq 16$, and
for\footnote{
We exclude $N=2$, since in this case one has to
calculate the norm of each Fock state explicitly
to avoid overlapping states, which 
is computationally very intensive.} 
$N=3,10,100$ and $1000$. Strictly speaking, there is no upper
limit on the size of $N$, since it appears as an algebraic
variable in the DLCQ matrix. 
The mass $m$ of the fermion (see equation (\ref{pminus}))
is set to zero. The results we obtain for $N=1000$ agree with the 
large $N$ results presented in \cite{bdk93} to at least
six significant figures.   

As we stated earlier, our main objective is to determine 
whether the multi-particle states of the $N=\infty$ spectrum
also persist at any finite $N$.
Towards this end, we first calculate the 
masses of various ``single particle''
states -- namely, the 
two lightest fermion and boson single-particle states -- for 
various values
of $K$ and $N$. The results\footnote{
We have also identified a number of other single
particle states but they are not relevant for the discussion presented here.}
are presented in 
Table \ref{fermionsingle} and Table \ref{bosonsingle}.

Note that for small values of $K$ there is no dependence
on $N$ (after expressing $M^2$ in units $g^2 N/\pi$).
Moreover, for larger values of $K$, the $N$ dependence
is surprisingly small. This seems puzzling at first, since we 
know interactions between single and multi-trace states
are governed by the `string coupling' $1/N$, and
so for $N=3$, one expects considerable multi-trace contributions
in a state such as $|F2\rangle$, which turns out to be
a superposition of mainly five-parton single trace Fock states.
Evidently
a rather remarkable cancellation
seems to be responsible for keeping the masses relatively 
independent of $N$. In this case, the limit $N=\infty$  provides an
excellent approximation to the $N=3$ case. 

%
\begin{table}[h!]
\begin{center}
\begin{small}
\begin{tabular}{||c|c|c|c|c||c|c|c|c|c||}
\hline 
\multicolumn{10}{| c |}{Fermion 
``Single Particle'' Masses (in units $g^2 N/\pi$)}\\
\hline
\multicolumn{5}{|| c ||}
{$|\mbox{F1} \rangle$} & 
\multicolumn{5}{c||}
  { $|\mbox{F2} \rangle$} \\
\hline $K$ & $N=3$  & $N=10$  & $N=100$  & $N=1000$  & 
       $K$ & $N=3$  & $N=10$  & $N=100$  & $N=1000$  \\
\hline
$3$ & 4.5 & 4.5 & 4.5 & 4.5 & 3 & - & - & - & - \\
\hline
5    & 5 & 5 & 5 & 5 & 5 & 12.5 & 12.5 & 12.5 & 12.5 \\
\hline
7  & 5.22272 & 5.22272 & 5.22272 & 5.22272 & 
7 & 14  & 14 & 14 &  14 \\
\hline
9  & 5.34559 & 5.34559  & 5.34559  & 5.34559  & 
9 &  14.7645  & 14.7645   & 14.7645  &  14.7645 \\
\hline
11 & 5.42226 & 5.42224 & 5.42224 & 5.42224 &
11 & 15.2575 & 15.2575 & 15.2575 & 15.2575 \\ 
\hline       
13 & 5.47410 & 5.47406 & 5.47406 & 5.47406 &
13 & 15.5909 & 15.5908 & 15.5908 & 15.5908 \\
\hline
15 & 5.51121 &  5.51115 & 5.51115 & 5.51114 & 
15 & 15.8314 & 15.8311  & 15.8311 & 15.8311 \\
\hline
\end{tabular}
\caption{The masses $M^2$ (in units $g^2 N/\pi$) 
of the two lightest ``single particle'' fermion
states, $|F1\rangle$ and $|F2\rangle$ respectively.
\label{fermionsingle}}
\end{small}
\end{center}
\end{table}
%
\begin{table}[h!]
\begin{center}
\begin{small}
\begin{tabular}{||c|c|c|c|c||c|c|c|c|c||}
\hline 
\multicolumn{10}{| c |}{Boson 
``Single Particle'' Masses (in units $g^2 N/\pi$)}\\
\hline
\multicolumn{5}{|| c ||}
{$|\mbox{B1} \rangle$}  & 
\multicolumn{5}{c||}
  { $|\mbox{B2} \rangle$} \\
\hline $K$ & $N=3$  & $N=10$  & $N=100$  & $N=1000$  & 
       $K$ & $N=3$  & $N=10$  & $N=100$  & $N=1000$  \\
\hline
$4$ & 8 & 8 & 8 & 8 & 4 & - & - & - & - \\
\hline
6    & 9 & 9 & 9 & 9 & 9 & 18 & 18 & 18 & 18 \\
\hline
8  & 9.49097  & 9.49097   & 9.49097  &  9.49097 & 
8 &  20  & 20 & 20 &  20 \\
\hline
10  & 9.78145  & 9.78145  & 9.78145  & 9.78145  & 
10 &   21.2117  & 21.2117   & 21.2117  & 21.2117 \\
\hline
12 & 9.97108  &  9.97103 &  9.97102 &  9.97102 &
12 & 22.0078 &  22.0078 &  22.0078 &  22.0078 \\ 
\hline       
14 & 10.1035 & 10.1034 & 10.1034  & 10.1034 &
14 & 22.5681  & 22.5680  & 22.5680  & 22.5680  \\
\hline
16 & 10.2006 & 10.2004 &  10.2004 & 10.2004 & 
16 & 22.9812 & 22.9811 &  22.9811 & 22.9811\\
\hline
\end{tabular}
\caption{The masses $M^2$ (in units $g^2 N/\pi$) 
of the two lightest ``single particle'' boson
states, $|B1\rangle$ and $|B2\rangle$ respectively.
  \label{bosonsingle}}
\end{small}
\end{center}
\end{table}

A crucial observation made by GHK \cite{ghk97} in their
recent work suggested that many of the remaining states
in the spectrum are adequately 
described as two or more freely interacting single 
particles: $|F1 \rangle \otimes |F1 \rangle$,
$|F1 \rangle \otimes |F2\rangle$,
$|F1 \rangle \otimes |B1\rangle$, and so on.
This is why the states in Tables \ref{fermionsingle} and 
\ref{bosonsingle} are referred to as ``single particles'', since
they themselves are not seen to decompose into more
fundamental stable particles.

The validity of this scheme may be tested in the DLCQ
framework as follows \cite{ghk97}. For a given
DLCQ resolution $K$, the mass $M_{F1 \otimes F1}(K)$
of the composite state
$|F1 \rangle \otimes |F1 \rangle$ is given by 
free body kinematics according to the relation 
\begin{equation}
\label{DLCQfreebody}
 \frac{M_{F1\otimes F1}^2(K)}{K} = 
\frac{M_{F1}^2(n)}{n} +\frac{M_{F1}^2(K-n)}{K-n}, 
\end{equation}
where $n$ is any positive integer less than $K$,
and where $M_{F1}(n)$ and $M_{F1}(K-n)$ are masses
of the $|F1 \rangle$ particle carrying $n$ and $K-n$ momentum 
units respectively.
In Table \ref{masspredictions},
we have applied equation (\ref{DLCQfreebody})
to determine the mass of 
$|F1\rangle \otimes |F1\rangle$ for various choices
of $n$ and $K$.
Since the mass of $|F1\rangle$ for each $K$ is essentially independent
of $N$ at five significant figures, 
the masses given in Table \ref{masspredictions}
are applicable for any $N$.
\begin{table}[h!]
\begin{center}
\begin{tabular}{|c|c|c|c|}
\hline
\multicolumn{4}{|c|}{Calculated Masses for 
$|F1 \rangle \otimes |F1 \rangle$ } \\
\hline
$K$ & 
\multicolumn{3}{|c|}{ $M^2_{F1 \otimes F1}(K)$}  \\
\hline
8 & 20 & - & - \\
\hline
10 & (20.00) & 22.461 & - \\
\hline 
12 & 20.953 & 25.127 & -  \\
\hline
14 & (20.891) & 22.315 & 27.901  \\
\hline    
16 & 21.441 & 23.887 & 30.737   \\
\hline 
\end{tabular}
\caption{Predicted values for  
$M^2$  (in units $g^2 N/\pi$) for the composite particle
$|F1 \rangle \otimes |F1 \rangle$ 
according to the two-free-body formula (\ref{DLCQfreebody}),
for various choices of $n$ and $K$. The different 
numbers appearing in a given row correspond to varying the value of $n$
in equation (\ref{DLCQfreebody}). Numbers that are in parentheses
correspond to a pair of identical $|F1 \rangle$ particles
(i.e. carrying the same momentum), and are therefore expected to 
be absent from the spectrum because of Fermi statistics. 
 \label{masspredictions}}
\end{center}
\end{table}
In Table \ref{actualbosonmasses} 
we list the actual masses for bosons 
that are observed in the DLCQ spectrum for various choices
of $K$ and $N$.  
\begin{table}[h!]
\begin{center}
\begin{small}
\begin{tabular}{|c|c|c|c|c|c|c|c|}
\hline 
\multicolumn{8}{| c |}{Observed Boson 
Masses $M^2$ in DLCQ Spectrum }\\
\hline
 $K$ & $N$ & \multicolumn{6}{|c|}{$M^2$ (in units $g^2 N/\pi$)} \\
\hline
8 & 3 &  $18.7312$ &  $20^{(2)}$ & 
-  & - & - & -  \\
\hline
8 & 1000 &  $18.7312$ &  $20^{(2)}$ & - & - & - & -   \\
\hline
10 & 3 & - &  - & 21.1213 & $22.461^{(2)}$ & - & -   \\
\hline
10 & 1000 & - & - & 21.1213 &  
$22.461^{(2)}$ & - & - \\
\hline
12 & 3 &  20.5603 & $20.9917^{(2)}$ &  
23.954 &
 $25.0849^{(2)}$ & - & -  \\
\hline
12 & 1000 &  
20.5048 & $20.9532^{(2)}$ &  24.0033 &
$25.1275^{(2)}$ & - & -  \\
\hline
14 & 3 & - & - &  21.7082 & $22.3756^{(2)}$ & 
26.8465 &
$27.8003^{(2)}$  \\
\hline
14 & 1000 & - & - &  21.6346 & $22.3154^{(2)}$ & 
 26.9517 & $27.9010^{(2)}$  \\
\hline
16 & 3 &  21.2700 & $21.4620^{(2)}$ & 
 23.2368 & $23.9392^{(2)}$ & 29.7486 &
$30.5640^{(2)}$  \\
\hline
16 & 1000 &  21.2303 & $21.4409^{(2)}$  &
 23.1910 & $23.8869^{(2)}$ & 29.9211 & $30.7373^{(2)}$  \\
\hline
\end{tabular}
\caption{Actual values for the mass squared $M^2$ (in units $g^2 N/\pi$)
of several bosonic states for different $N$ and $K$ (excluding
single particle states). 
Numbers with a superscript ${}^{(n)}$ correspond
to an exact $n$-fold degeneracy in the spectrum.    
\label{actualbosonmasses}}
\end{small}
\end{center}
\end{table}
Comparing Tables \ref{masspredictions} and \ref{actualbosonmasses},
we see for $N=1000$ (i.e. essentially `large $N$'),
that the masses corresponding to two free $|F1\rangle$ particles
appear in the actual DLCQ spectrum with an exact two-fold degeneracy.
In addition, the masses in Table \ref{masspredictions} that
correspond to two identical $|F1\rangle$ 
particles (in parentheses) are absent from the 
DLCQ spectrum, which is of course consistent
from the Fermi statistics of two  freely
interacting identical fermions. 
The two-fold degeneracy we see here in the bosonic spectrum does not 
occur in the $N=\infty$ analysis of GHK \cite{ghk97},
and this can be easily understood by analyzing the Fock state
content of each doublet. What we find for any doublet at $N=1000$
is that one state is a superposition of essentially single-trace
Fock states, while the other is mainly a superposition
of two-trace Fock states. Since the Hilbert space
for the $N =\infty$ theory is generated from single-trace
Fock states only, the multi-trace states we see here
will be absent in the $N=\infty$ spectrum.

One remarkable property of the theory, which can be observed
from Table \ref{actualbosonmasses}, is that for 
small values of $N$ -- say, $N=3$ (where
$1/N$ interactions can no longer be neglected) -- the exact two-fold
degeneracy still survives, and the actual masses deviate only
slightly (if at all) from the large $N$ results. 
Nevertheless,
the Fock state content of each state in a doublet 
is drastically altered if we vary $N$;
single-trace
and multi-trace Fock states now contribute 
equally in each state for small $N$. Evidently, significant
cancellations between $1/N$ contributions in the finite $N$ Hamiltonian
must be occurring in order to keep masses relatively 
independent of $N$. We should point out that the
additional term in equation (\ref{pminus}) for $P^-$
that distinguishes it from the $N=\infty$ Hamiltonian
is crucial in these calculations -- no such mass degeneracy 
would be observed if it was omitted.
It is still unclear at this point whether the very small
deviations from the predicted masses in 
Table \ref{masspredictions} and the mass of the 
doublets appearing
in the finite $N$ DLCQ
spectrum in Table \ref{actualbosonmasses} will remain in
the continuum $K \rightarrow \infty$ limit. There is already
evidence in Table \ref{actualbosonmasses} suggesting that the 
deviations are diminishing for larger values of $K$, but 
clarifying this issue will certainly require solving
the DLCQ spectrum for larger values of $K$, and we leave this for
future work.

As was remarked in the work of GHK \cite{ghk97}, the remaining
boson states listed in Table \ref{actualbosonmasses} 
agree with the masses presented in Table
\ref{masspredictions} up to $1/K$; i.e. as we increase the
harmonic resolution, the agreement becomes more exact.
These states are therefore multi-particle states that
give rise to two-body continua in the continuum limit 
$K \rightarrow \infty$. 
We also note that this picture is not disturbed for relatively
small values of $N$ (such as $N=3$), and so the observations made 
by GHK in the context of the $N=\infty$ theory also appear to be
valid at any finite $N$. In particular, for any finite $N$,
our results are consistent with the interpretation that  
the spectrum of the (continuum) theory  
becomes continuous above the threshold mass
 $M^2 = 4M_{F1}^2$, where $M_{F1}$ is the 
mass of the lightest stable particle
in the continuum theory.  

\medskip

An identical analysis may be performed for
the composite state $|F1 \rangle \otimes |F2 \rangle$, and
we find once again that the masses predicted by 
equation (\ref{DLCQfreebody}) emerge in the actual DLCQ spectrum
as mass doublets for any finite $N$. 
Agreement is more precise for large values of $N$, but we nevertheless
obtain good accuracy even for $N=3$ as we did for 
$|F1 \rangle \otimes |F1 \rangle$ in Table \ref{actualbosonmasses}. 
For composite
particles with fermion statistics, such as 
$|F1\rangle \otimes |B1\rangle$, the masses predicted by 
equation (\ref{DLCQfreebody}) also emerge in the DLCQ spectrum
for any $N$, but we no longer see the exact two-fold degeneracy
that was observed in the bosonic sector. The fermion states
that exactly satisfy equation (\ref{DLCQfreebody})
for large $N$ 
are not observed in the $N=\infty$ analysis of GHK,
since these states are a superposition
of two-trace Fock states (one fermion, and one boson)
for large values of $N$. For $N=3$, such states
involve a superposition of both single and multi-trace states.    

\medskip

As was pointed out in \cite{ghk97}, a large
portion of the ``multi-particle''
spectrum in the fermionic sector can be summarized surprisingly well
if we replace equation (\ref{DLCQfreebody}) with
the modified formula 
\begin{equation}
\label{modifiedfreebody}
\frac{M^2}{K-1} = \frac{M^2_{F1}(n)}{n} +
                    \frac{M^2_{F1}(K-n-1)}{K-n-1}.
\end{equation} 
In other words, some fermionic states at resolution $K$ in the 
spectrum appear to be well approximated as two $|F1\rangle$
particles  with a combined momentum of $K-1$. There is still
one unit of momentum unaccounted for, and one proposal \cite{ghk97}
is to assume that this momentum is associated with
a topological configuration of the light-cone vacuum that makes
it distinct from the trivial light-cone vacuum.
In fact, in order to get a fermion state overall, we
need to assume that this topological vacuum carries
fermion quantum numbers.

Although it is indeed possible to construct topologically
distinct vacuum sectors in an SU($N$) light-cone gauge theory
with the properties cited above \cite{pin97a,mrp97},
the empirical success of equation (\ref{modifiedfreebody})
lends itself to an alternative interpretation.
Firstly, we know the theory has a supersymmetric 
point for a very specific value of the fermion mass \cite{ku93}: 
$m^2 = g^2 N/\pi$.   As we reduce the fermion mass $m$
to zero, one can show numerically
that the mass degeneracy 
between the lightest boson-fermion supersymmetric pair
is broken linearly as a function
of $m^2/g^2 N$ \cite{ku93}. However, if we consider a heavier
supersymmetric pair, this behavior is not necessarily observed.
For example, at the supersymmetric point ($m^2 = g^2 N/\pi$), the 
mass of the next-to-lightest fermion in the $K=15$ DLCQ
spectrum is approximately $M^2 = 43.5$, while its super-partner
is identified in the $K=14$ DLCQ spectrum as a boson with
mass\footnote{The boson and fermion masses are expected
to converge together in the continuum limit $K \rightarrow \infty$.}  
$M^2 = 40.7$ (in units $g^2 N/\pi$). The boson state
is essentially a 
mixture of two-bit and four-bit single-trace states,
while the fermion is observed to be essentially a mixture of
three-bit single-trace states.

If we carefully track these states as the fermion
mass $m$ is decreased to zero, we find in the massless limit that the 
fermion state has mass $M^2 = 20.86$,
while its (former) superpartner has the new mass $M^2 = 21.63$. 
Surprisingly,
the masses are still approximately degenerate, so supersymmetry
appears to be very weakly broken for these  states.
This suggests that the theory is in fact {\em asymptotically
supersymmetric} (i.e. the spectrum of highly excited 
bound states is almost supersymmetric for any fermion mass $m$),
an observation  pointed out by Boorstein and Kutasov \cite{kub97}.
One can use formula (\ref{DLCQfreebody}) to show that
the boson is in fact a composite state of two freely
interacting $|F1\rangle$ particles \cite{ghk97}. We
conclude therefore that the mass of particular fermionic states
at DLCQ resolution $K$ may be estimated by an application of
equation (\ref{DLCQfreebody}) with $K$ replaced with $K-1$,
since the composite boson at resolution $K-1$ is still approximately
equal in mass with the fermion state at resolution $K$.   
This scheme provides a natural derivation of equation 
(\ref{modifiedfreebody}) from (\ref{DLCQfreebody}).

\medskip

At this point, we have made no reference to composite particles made
from three or more single-particle states; to further validate 
our interpretation of the spectrum, we 
look for states of the form $|F1\rangle \otimes |F1\rangle \otimes
 |F1\rangle$. The mass $M^2_{F1\otimes F1 \otimes F1}$ is now
given by a simple generalization of equation (\ref{DLCQfreebody}):
\begin{equation}
\frac{M^2_{F1\otimes F1 \otimes F1}}{K} =
  \frac{M^2_{F1}(K_1)}{K_1} +  \frac{M^2_{F1}(K_2)}{K_2}
 + \frac{M^2_{F1}(K_3)}{K_3},
\end{equation}
where $K_1+K_2+K_3 = K$. For example, setting $K_1=3, K_2=5$, and
$K_3=7$ gives $M^2_{F1\otimes F1 \otimes F1} =  48.6915$. This
is indeed observed in the DLCQ fermionic
spectrum at $K=15$, and we observe an exact two-fold degeneracy
at any finite $N$. These states involve significant contributions
from multi-trace Fock states, and are therefore only seen in
a finite $N$ analysis.

\section{Discussion}
\label{conclusions}
In this work we have implemented DLCQ to study the
spectrum of $\mbox{QCD}_{1+1}$ coupled to a massless
Majorana fermion transforming
in the adjoint representation of the gauge
group SU($N$), where  $N$ is any integer $N\geq 3$. We observed
in Tables \ref{fermionsingle} and \ref{bosonsingle}
that the masses of the stable ``single-particle'' states
in the spectrum are very weakly dependent on $N$.
Evidently, considerable
cancellations must be responsible for
protecting masses from large $1/N$ corrections. 
It would be desirable to have a better 
understanding of this.

After comparing predicted masses for
the 
composite particle $|F1\rangle \otimes |F1 \rangle$
(Table \ref{masspredictions}) 
with the actual DLCQ spectrum for bosons 
(Table \ref{actualbosonmasses}), we observed that
the finite $N$ spectrum is consistently described
as multi-particles
with freely interacting stable constituents.    
Therefore, the existence of a continuous spectrum
above the threshold mass $M^2 = 4 M_{F1}^2$
appears to be valid not only for $N=\infty$ \cite{ghk97}, but also
for any finite $N$. This behavior is compatible
with the viewpoint that the finite $N$ theory
is in a  screening phase \cite{gkm96}.
Clearly, probing larger values of $K$ will help
clarify this issue, and we leave this challenge for future work.

We also attempted to explain the fermionic
multi-particle spectrum by assuming the 
theory was asymptotically supersymmetric; i.e.
excited states in the spectrum are almost
supersymmetric for any fermion mass, including $m=0$.
An explicit example was given.
This enables one to connect the mass of a
(sufficiently heavy) fermion
with its corresponding bosonic
`super-partner'. 
Because the DLCQ resolution of a boson and
fermion state differ by one unit of momentum 
(after imposing anti-periodic
boundary conditions for fermions), we were able to 
derive naturally formula (\ref{modifiedfreebody})
from equation (\ref{DLCQfreebody}) describing the mass
of two free particles. 

Finally, we comment that the finite $N$ analysis we have adopted
here affords interesting insights into
large $N$ theories in general; solving a theory for 
increasing values of $N$ enables one to study the
large $N$ limit in a way that is not accessible in 
the $N=\infty$ -- or planar -- approximation. In particular,
subtleties can arise in the large $N$ limit where 
single and multi-trace states co-exist on an equal
footing. This is especially true for theories that
may be susceptible to a phase transition -- such
as the model studied here -- and may play a crucial
role in understanding the fundamental degrees of freedom
of large $N$ gauge theories.  

\medskip
\begin{large}
 {\bf Acknowledgments}
\end{large}

The authors would like to thank Igor Klebanov and
Oleg Lunin for many useful discussions.
F.A. is gratfeul to Jungil Lee for assistance with computer work.

\vfil

\end{document}